\documentclass[twocolumn]{aastex631}
\usepackage{graphicx,natbib,bm,url,color}
\graphicspath{{./fig/}{./png/}}

\newcommand{\EQ}{\begin{equation}}
\newcommand{\EN}{\end{equation}}
\newcommand{\EQA}{\begin{eqnarray}}
\newcommand{\ENA}{\end{eqnarray}}

\newcommand{\Eq}[1]{Eq.~(\ref{#1})}

\newcommand{\Sec}[1]{Sect.~\ref{#1}}

\newcommand{\Fig}[1]{Figure~\ref{#1}}

\newcommand{\Figs}[2]{Figures~\ref{#1} and \ref{#2}}

\newcommand{\Tab}[1]{Table~\ref{#1}}

\newcommand{\bra}[1]{\langle #1\rangle}

\newcommand{\meanrho}{\overline{\rho}}

{}
{}

\newcommand{\meanSSSS}{\overline{\mbox{\boldmath ${\mathsf S}$}} {}}

{}
{}
{}
{}
{}
{}
{}
{}
\newcommand{\meanEE}{\overline{\mbox{\boldmath $E$}}{}}{}
{}
{}
{}
{}
{}
{}
{}
{}
\newcommand{\meanUU}{\overline{\bm{U}}}
\newcommand{\meanVV}{\overline{\bm{V}}}

{}

\newcommand{\meanA}{\overline{A}}
\newcommand{\meanB}{\overline{B}}

\newcommand{\meanU}{\overline{U}}

\newcommand{\meanp}{\overline{p}}

{}

{}
{}
{}

\newcommand{\xxx}{\hat{\mbox{\boldmath $x$}} {}}
\newcommand{\yyy}{\hat{\mbox{\boldmath $y$}} {}}

\newcommand{\meanAA}{{\overline{\bm{A}}}}
\newcommand{\meanBB}{{\overline{\bm{B}}}}
\newcommand{\meanJJ}{{\overline{\bm{J}}}}

\newcommand{\BB}{\bm{B}}

\def\oA{\omega_{\rm A}}

\newcommand{\nab}{{\bm{\nabla}}}

\newcommand{\OO}{\bm{\Omega}}

\newcommand{\ii}{{\rm i}}

\newcommand{\DD}{{\rm D} {}}

\newcommand{\dd}{{\rm d} {}}

\newcommand{\const}{{\rm const}  {}}

\def\la{\mathrel{\mathchoice {\vcenter{\offinterlineskip\halign{\hfil
$\displaystyle##$\hfil\cr<\cr\sim\cr}}}
{\vcenter{\offinterlineskip\halign{\hfil$\textstyle##$\hfil\cr<\cr\sim\cr}}}
{\vcenter{\offinterlineskip\halign{\hfil$\scriptstyle##$\hfil\cr<\cr\sim\cr}}}
{\vcenter{\offinterlineskip\halign{\hfil$\scriptscriptstyle##$\hfil\cr<\cr\sim\cr}}}}}

\def\H{\mbox{\rm H}}

\def\Ta{\mbox{\rm Ta}}

\def\Pm{\mbox{\rm Pr}_{\rm M}}

\def\EEK{{\cal E}_{\rm K}}
\def\EEM{{\cal E}_{\rm M}}
\def\EEMeq{{\cal E}_{\rm M}^\mathrm{eq}}
\def\EEMX{{\cal E}_{\rm M}^X}
\def\EEMZ{{\cal E}_{\rm M}^Z}

\def\cs{c_{\rm s}}
\def\Cs{C_{\rm s}}

\def\vA{v_{\rm A}}

\def\vAz{v_{\rm A0}}
\def\vAeq{v_{\rm A}^{\rm eq}}

\def\epsK{\epsilon_{\rm K}}
\def\epsM{\epsilon_{\rm M}}

\def\Brms{B_{\rm rms}}
\def\meanBrms{\overline{B}_{\rm rms}}

\def\urms{u_{\rm rms}}

\def\nuT{\nu_{\rm T}}

\def\etaT{\eta_{\rm T}}

\def\Beq{B_{\rm eq}}

\newcommand{\W}{\,{\rm W}}
\newcommand{\V}{\,{\rm V}}

\newcommand{\T}{\,{\rm T}}
\newcommand{\G}{\,{\rm G}}

\newcommand{\K}{\,{\rm K}}
\newcommand{\g}{\,{\rm g}}
\newcommand{\s}{\,{\rm s}}

\newcommand{\cm}{\,{\rm cm}}

\newcommand{\m}{\,{\rm m}}

\newcommand{\Mm}{\,{\rm Mm}}

\newcommand{\J}{\,{\rm J}}

\newcommand{\A}{\,{\rm A}}

\def\apjl{ApJL}
\def\apjs{ApJS}

\def\aap{A\&A}

\begin{document}

\title{Magnetorotational instability in a solar near-surface mean-field dynamo}

\author[0000-0002-7304-021X]{Axel Brandenburg}
\affiliation{Nordita, KTH Royal Institute of Technology and Stockholm University, Hannes Alfv\'ens v\"ag 12, SE-10691 Stockholm, Sweden}
\affiliation{The Oskar Klein Centre, Department of Astronomy, Stockholm University, AlbaNova, SE-10691 Stockholm, Sweden}
\affiliation{McWilliams Center for Cosmology \& Department of Physics, Carnegie Mellon University, Pittsburgh, PA 15213, USA}
\affiliation{School of Natural Sciences and Medicine, Ilia State University, 3-5 Cholokashvili Avenue, 0194 Tbilisi, Georgia}

\author[0009-0008-4918-3852]{Gustav Larsson}
\affiliation{Nordita, KTH Royal Institute of Technology and Stockholm University, Hannes Alfv\'ens v\"ag 12, SE-10691 Stockholm, Sweden}
\affiliation{Department of Physics, Stockholm University, AlbaNova, 10691 Stockholm, Sweden}

\author[0000-0001-9268-4849]{Fabio Del Sordo}
\affiliation{Scuola Normale Superiore, Piazza dei Cavalieri, 7 56126 Pisa, Italy}
\affiliation{INAF, Osservatorio Astrofisico di Catania, via Santa Sofia, 78 Catania, Italy}

\author[0000-0001-9619-0053]{Petri J. K\"apyl\"a}
\affiliation{Institut f\"ur Sonnenphysik (KIS), Georges-K\"ohler-Allee 401a, 79110 Freiburg im Breisgau, Germany}

\begin{abstract}
We address the question whether the magnetorotational instability
(MRI) can operate in the near-surface shear layer (NSSL) of the Sun and
how it affects the interaction with the dynamo process.
Using hydromagnetic mean-field simulations of $\alpha\Omega$-type dynamos
in rotating shearing-periodic boxes, we show that for negative shear,
the MRI can operate above a certain critical shear parameter.
This parameter scales inversely with the equipartition magnetic field
strength above which $\alpha$ quenching set in.
Like the usual $\Omega$ effect, the MRI produces toroidal magnetic field
when the field is sufficiently strong.
The work done by the Lorentz force is positive, so the magnetic field
drives kinetic energy and not the other way around, as in a turbulent dynamo.
This results in strong kinetic energy production and dissipation,
which occurs at the expense of the magnetic energy.
In view of the application to the solar NSSL, we conclude that the
turbulent magnetic diffusivity may be too large for the MRI to be
excited and that therefore only the standard $\Omega$ effect is expected
to operate.
\end{abstract}

\section{Introduction}

The magnetorotational instability (MRI) provides a source of turbulence
in accretion discs, where it feeds on Keplerian shear to turn potential
energy into kinetic and magnetic energies; see \cite{BH98} for a review.
For the MRI to be excited, the angular velocity $\Omega$ must decrease
with increasing distance $\varpi$ from the rotation axis, i.e.,
$\partial\Omega/\partial\varpi<0$.
There must also be a moderately strong magnetic field.
This condition is obeyed not only in accretion discs, but
also in stars like the Sun, where both requirements may be satisfied in the near-surface
shear layer (NSSL), the outer 4\% of the solar radius \citep{Schou+98}.
This motivated \cite{Vasil+24} to study the excitation properties
of the MRI in the NSSL of the Sun using spherical global modes.

The possible relevance of the MRI for stellar radiative zones has been
discussed for a long time \citep{Balbus+Hawley94, Urpin96, Menou+04}.
\cite{Parfrey+Menou07} proposed that small-scale magnetorotational
turbulence prevents coherent magnetic dynamo action in the tachocline
at higher latitudes.
They argued that this could explain the latitudinal restriction of solar
active regions to the vicinity of the equator.
\cite{Masada11} suggested that the MRI could play a role in maintaining
thermal wind balance in the Sun.
\cite{Kagan+Wheeler14} found that nonaxisymmetric MRI modes tend to
grow faster in the Sun than the corresponding axisymmetric modes.
\cite{Wheeler+15} found the MRI to play a role in the later stages of
massive star evolution.
Unlike the recent work by \cite{Vasil+24}, these papers only considered
local analyses.
In addition to the Sun and other stars, proto-neutron stars represent
a particularly important application \citep{Reboul-Salze+22}.

In the Sun's outer 30\% by radius there is convection converting
part of the Sun's thermal energy into kinetic energy.
The nonuniform rotation of the Sun is explained by the fact that the
convection is anisotropic such that solid-body rotation is no longer a solution
to a rotating fluid even in the absence of external torques \citep{
Lebedinskii41, Wasiutynski46, Kippenhahn63, Kohler70, Ruediger80, Bra+90}.
This causes also the emergence of the aforementioned NSSL
\citep{Ruediger+14, Kitchatinov16, Kitchatinov23}.
In addition, small-scale \citep{MP89, Nordlund+92, Bra+96,
Cattaneo99} and large-scale \citep{KKB08, HP09, MS14, Bushby+08}
magnetic fields exist as a result of the convective turbulence.
The presence of radial stratification in density and/or turbulent
intensity, together with global rotation, causes the occurrence of
large-scale magnetic fields \citep{Mof78, Par79, KR80, ZRS83}.
Thus, in the Sun, the two ingredients of the MRI---differential rotation
and magnetic fields---are ultimately caused by the underlying
convection.

\cite{Vasil+24} argue that an initial poloidal magnetic field is only
present at the start of the cycle to get the MRI excited.
Furthermore, they argue that the MRI itself is important in driving
the dynamo process.
However, the magnetic field required to sustain the MRI to work must
come from a dynamo process and cannot rely on the initial field.
To address the question of whether or not the MRI is
excited and whether it contributes to shaping the Sun's magnetic field
to display equatorward migration of a global large-scale magnetic field,
we need to separate the MRI-driven flows from the convection.
One approach is to ignore convection but to retain some of its
secondary effects, i.e., the NSSL with $\partial\Omega/\partial\varpi<0$
and magnetic fields; see the discussion by
\cite{Vasil+24} and an appraisal by \cite{Zweibel24}.
Another approach, the one taken here, is to average over the convection.
By employing azimuthal averages, one is left with a stationary,
nonturbulent background.
Furthermore, correlations among different components of the fluctuating
parts of the turbulent velocity and magnetic fields emerge that are
parameterized in terms of (i) diffusive contributions, such as turbulent
viscosity and turbulent magnetic diffusion, and (ii) nondiffusive
contributions such as $\Lambda$ and $\alpha$ effects.
There is no universal agreement about the relevance of these effects
\citep[e.g.,][]{Spruit11, Hughes18}.
In the mean-field description of the Sun, they are crucial effects
able to explain the production of differential rotation and large-scale
magnetic fields \citep{RH04}.
These effects can explain the NSSL and the large-scale magnetic field by
solving the averaged form of the underlying equations \citep{Pipin17};
see \cite{Bra+23} for a review.

Apart from the $\alpha$ effect, there are also several other mean-field
effects that can produce large-scale magnetic fields: turbulent pumping
with a time delay, negative turbulent magnetic diffusivity, the R\"adler
effect, the shear--current effect, and the incoherent $\alpha$--shear effect;
see Table~2 of \cite{BN23}.
In addition, in phenomenological studies of solar magnetism, the
decay of active regions is sometimes associated with a dynamo effect
\citep{Leighton69}.
In its simplest form, however, it is just an $\alpha$ effect \citep{Stix74}.
In a more advanced formulation, it can be written as a nonlocal $\alpha$
effect; see \cite{DC99} for an example implementation in a mean-field model
and \cite{BK07} for results regarding the effect on magnetic helicity
conservation in the Sun.
In any case, it would not be possible to adopt such an effect here due
to the top--bottom symmetry of our model.

In principle, it is possible to study the interaction between the
MRI and the dynamo in fully three-dimensional turbulence simulations.
However, the essentials of these processes may well be captured in
a mean-field approach.
Using direct numerical simulations with forced turbulence,
\cite{Vaisala+14} demonstrated that the onset of the MRI is
consistent with what is expected from mean-field estimates.
In particular, the onset requires larger magnetic Reynolds numbers 
than in the ideal case due to the action of turbulent diffusion.

Averaging over the convective motions of the Sun has been done previously
in the context of mean-field hydrodynamics with the $\Lambda$ effect.
When including compressibility and thermodynamics, it was noticed that
the equations display an instability \citep{Gierasch+74, Schmidt+82,
Chan+87, RT91, RS92}, whose nature was not initially understood.
However, this later turned out to be an example where averaging over
the convection leads to mean-field equations that themselves are
susceptible to an instability, namely the onset of convection.
This depends on how close the mean-field state is to adiabatic and on
the values of the turbulent viscosity and turbulent thermal diffusivities
are \citep{Tuominen+94}.

When magnetic fields are present and sustained by a dynamo, the full
system of magnetohydrodynamic (MHD) equations may be unstable to the MRI.
We must emphasize that we are here not talking about the
previously studied case where the MRI provides the source of turbulence,
which then reinforces an initial magnetic field by dynamo action through
a self-sustained doubly positive feedback cycle \citep{BNST95, Hawley+96,
Stone+96}.
Even in that case, a mean-field description may be appropriate to
quantify the nature of a large-scale dynamo governed by rotation and
stratification \citep{Bra+Sok02, Bra05, Gressel+10}.
However, such a description can only be an approximate one, because
the level of turbulence is unknown and emerges only when solving the
underlying, essentially nonlinear dynamo problem \citep{Rincon+07,
Lesur+Ogilvie08, Herault+11}.

In the present paper, we focus on the simpler case where a mean-field
dynamo is assumed to exist but may be modified by the MRI.
Ideally, in view of solar applications, it would be appropriate to
consider an axisymmetric hydromagnetic mean-field dynamo with differential
rotation being sustained by the $\Lambda$ effect.
Such systems have been studied for a long time \citep{Bra+90, Bra+91,
Bra+92, Kitchatinov+Ruediger95, Rempel06, Pipin17, Pipin+Kosovichev19},
but no MRI was ever reported in such studies.
One reason for this might be that it is hard to identify the operation
of the MRI in a system that is already governed by a strong instability
which is responsible for producing the magnetic field.
We therefore take a step back and consider here a system in Cartesian
geometry.
In Section 2, we provide the details of our model, and present the
results in Section 3.
We conclude in Section 4.

\section{Our model}

\subsection{Shearing box setup}

Following the early work of \cite{BH91, BH92} and \cite{HB91, HB92}, we
study the MRI in a shearing-periodic box, where $x$ is the cross-stream
direction, $y$ is the streamwise or azimuthal direction, and $z$ is the
spanwise or vertical direction.
As in \cite{Vaisala+14}, we consider the mean-field equations for
azimuthally averaged velocities $\meanUU(x,z,t)$, the magnetic field
$\meanBB(x,z,t)$, and the mean density $\meanrho(x,z,t)$.
The system is rotating with the angular velocity $\OO$, and there is
a uniform shear flow $\meanVV(x)=(0, Sx, 0)$, so the full velocity is
therefore given by $\meanVV+\meanUU$.
We consider the system to be isothermal with constant sound
speed $\cs$, so the mean pressure $\meanp(x,z,t)$ is given by
$\meanp=\meanrho\cs^2$.
The mean magnetic field is expressed in terms of the mean magnetic
vector potential $\meanAA(x,z,t)$ with $\meanBB=\nab\times\meanAA$
to satisfy $\nab\cdot\meanBB=0$.
The full system of equations for $\meanrho$, $\meanUU$, and $\meanAA$
is given by \citep{BNST95, BRRK08}
\begin{equation}
\frac{\DD\ln\meanrho}{\DD t}=-\nab\cdot\meanUU
\label{dUdt}
\end{equation}
\begin{eqnarray}
\frac{\DD\meanUU}{\DD t}=&-&S\meanU_x\yyy-2\OO\times\meanUU-\cs^2\nab\ln\meanrho \nonumber \\
&+&\left[\meanJJ\times\meanBB+\nab\cdot(2\nuT\meanrho\meanSSSS)\right]/\meanrho,
\label{dUdt}
\end{eqnarray}
\begin{equation}
\frac{\partial\meanAA}{\partial t}=-S\meanA_y\xxx+\meanUU\times\meanBB+\alpha\meanBB-\etaT\mu_0\meanJJ,
\label{dUdt}
\end{equation}
where $\DD/\DD t=\partial/\partial t+\meanUU\cdot\nab$ is the advective derivative,
$\meanSSSS$ is the rate-of-strain tensor of the mean flow with the
components $\overline{\mathsf{S}}_{ij}=(\partial_i \meanU_j+\partial_j
\meanU_i)/2 -\delta_{ij}\nab\bm\cdot\meanUU/3$, $\OO$ is the angular velocity,
$S=-q\Omega$ is the shear parameter,
and $\meanJJ=\nab\times\meanBB/\mu_0$ is the mean current density
with $\mu_0$ being the vacuum permeability.
There are three mean-field parameters: the turbulent viscosity $\nuT$,
the turbulent magnetic diffusivity $\etaT$, and the $\alpha$ effect.
Note that in our two-dimensional case, $\meanVV\cdot\nab=Sx\partial_y=0$.
In some cases, we allow for $\alpha$ quenching and write
\begin{equation}
\alpha=\alpha_0/(1+\meanBB^2/\Beq^2),
\end{equation}
where $\Beq$ is the equipartition field strength above which $\alpha$
begins to be affected by the feedback from the Lorentz force of the
small-scale magnetic field \citep{IR77}.
We sometimes refer to this as microphysical feedback to distinguish it
from the macrophysical feedback from the Lorentz force of the
large-scale magnetic field, $\meanJJ\times\meanBB$.
This type of saturation is sometimes also called the Malkus--Proctor
mechanism, after the early paper by \cite{Malkus+Proctor75}, who employed
spherical geometry.

In the absence of $\alpha$ quenching ($\Beq\to\infty$), the only
possibility for the dynamo to saturate is via the Lorentz force from the
mean magnetic field, $\meanJJ\times\meanBB$, i.e., the aforementioned Malkus--Proctor
mechanism.
Also relevant to our present work is that of \cite{Schuessler79}, who
considered Cartesian geometry.
Our solutions, however, are simpler still in that we employ periodic
boundary conditions in most cases.

A simple way to identify the operation of the MRI in a dynamo is by comparing models with positive and
negative values of $q$, because the MRI only works in the range $0<q<2$.
Note also that for $q>2$, the hydrodynamic state is Rayleigh unstable
and results in an exponentially growing shear flow, $\meanU_y(z)$,
without ever saturating in a periodic system.
For the solar NSSL, however, we have $q=1$ \citep{Barekat+14}.

Given that our main interest lies in the investigation
of the effect of the MRI on a dynamo, where any value of $q$
in the range $0<q<2$ is of interest, we have chosen here
$q=3/2$ for Keplerian accretion disks.
Smaller values of $q$ reduce the stress by a factor $q/(2-q)$
\citep{Abramowicz+96}, but the MRI is qualitatively unchanged.
Below, we demonstrate with a few runs that this is true for most
quantities.
Obviously, for proper predictions for the Sun and sun-like stars,
not only $q=1$, but also proper spherical geometry must be used.
Finally, as motivated above, we also consider negative values of $q$.

Some of our models with positive shear ($S>0$ or $q<0$), where the MRI is inactive,
do not saturate in the absence of $\alpha$ quenching.
To check whether this is a peculiarity of the use of periodic boundary
conditions, we also consider models with what is called a vertical field
condition, i.e.,
\begin{equation}
\meanB_x= \meanB_y= \partial_z \meanB_z=0,
\end{equation}
which corresponds to $\partial_z\meanA_x=\partial_z\meanA_y=\meanA_z=0$.
Note that with this boundary condition the normal component
of the Poynting vector $\meanEE\times\meanBB/\mu_0$, where
$\meanEE=\etaT\mu_0\meanJJ-\meanUU\times\meanBB$ is the
mean electric field, vanishes.
Thus, energy conservation is still preserved.

\subsection{Input and output parameters}
\label{Input+OutputParameters}

We consider a two-dimensional domain $L_x\times L_z$ and define
$k_1=2\pi/L_z$ as our reference wavenumber, which is the lowest
wavenumber in the $z$-direction.
The lowest wavenumber in the $x$-direction is $k_{1x}=2\pi/L_x$.
Our main input parameters are
\begin{equation}
\label{eq:c_alpha-c_Omega}
C_\alpha=\alpha_0/\etaT k_1,\quad
C_\Omega=S/\etaT k_1^2,
\end{equation}
as well as $q=-S/\Omega$ and $\Beq$, which can be expressed via the
corresponding Alfv\'en speed, $\vAeq\equiv\Beq/\sqrt{\mu_0\rho_0}$,
in nondimensional form as
\begin{equation}
\mathcal{B}_\mathrm{eq}\equiv\vAeq k_1/\Omega.
\label{calBdef}
\end{equation}
In all our cases, we assume $\Pm\equiv\nuT/\etaT=1$ for the turbulent
magnetic Prandtl number.
The sound speed is specified in terms of the ratio $\Cs\equiv\cs k_1/\Omega$,
for which we take in most of the cases $\Cs=10$.
As we will see below, the kinetic energy of the generated flows is
typically well below $\rho_0\Omega^2/k_1^2$.
Therefore, the value of $\Cs=10$ is large enough so that the results are
not affected by compressibility effects---even in those cases where the
kinetic energy exceeds $\rho_0\Omega^2/k_1^2$ by some amount.

\begin{figure}\begin{center}
\includegraphics[width=\columnwidth]{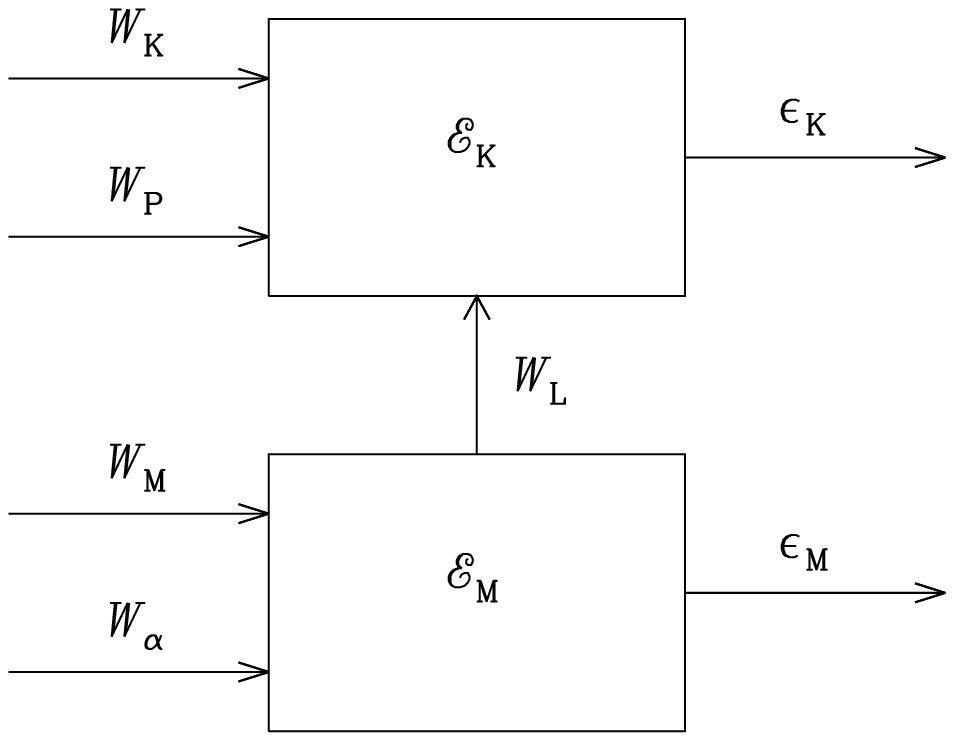}
\end{center}\caption{
Flow of energy in a hydromagnetic mean-field dynamo.
}\label{energetics}\end{figure}

The diagnostic output parameters are the energies of the mean fields that are
derived either under $yz$ or $xy$ averaging, $\EEM^{X}$ and $\EEM^{Z}$,
respectively.
These are sometimes normalized by $\EEMeq\equiv\Beq^2/2\mu_0$.
We also monitor various parameters governing the flow of energy in
our system.
These include the mean kinetic and magnetic energy densities,
$\EEK=\bra{\meanrho\meanUU^2/2}$ and $\EEM=\bra{\meanBB^2/2\mu_0}$,
their time derivatives, $\dot{\mathcal{E}}_\mathrm{K}$ and
$\dot{\mathcal{E}}_\mathrm{M}$,
the kinetic and magnetic energy dissipation rates,
$\epsK=\bra{2\meanrho\nuT\meanSSSS^2}$ and $\epsM=\bra{\etaT\mu_0\meanJJ^2}$,
the fluxes of kinetic and magnetic energy tapped from the shear flow,
$W_\mathrm{K}=\bra{\meanrho\meanU_x\meanU_y S}$ and
$W_\mathrm{M}=-\bra{\meanB_x\meanB_y S/\mu_0}$,
the work done by the pressure force, $W_\mathrm{P}=-\bra{\meanUU\cdot\nab\meanp}$
as well as the work done by the $\alpha$ effect,
$W_\alpha=\bra{\alpha\meanJJ\cdot\meanBB}$,
and the work done by the Lorentz force,
$W_\mathrm{L}=\bra{\meanUU\cdot(\meanJJ\times\meanBB)}$.
\Fig{energetics} gives a graphical illustration showing the flow
of energy in a hydromagnetic mean-field dynamo with shear.

For a uniform vertical magnetic field, $\BB_0=(0,0,B_0)$, the MRI is
excited when $\vAz k_1<\sqrt{2\Omega S}$, where $\vAz=B_0/\sqrt{\mu_0\rho_0}$
is the Alfv\'en speed of the uniform vertical magnetic field.
The MRI can be modeled in one dimension with $\nab=(0,0,\partial_z)$.
Such a one-dimensional setup could also lead to what is called an
$\alpha\Omega$ dynamo, which means that the mean radial or cross-stream
field $\meanB_x$ is regenerated by the $\alpha$ effect and the mean
toroidal or streamwise field $\meanB_y$ is regenerated by the $\Omega$
effect, or, more precisely, the shear flow $\meanVV(x)$.
One sometimes also talks about an $\alpha^2$ dynamo if there is no shear,
or about an $\alpha^2\Omega$ dynamo if both the $\alpha$ effect and shear
contribute to regenerating $\meanB_y$.

In the one-dimensional case with $\nab=(0,0,\partial_z)$ and periodic boundary
conditions, the $\alpha^2$ dynamo is excited when $C_\alpha>1$, while
the $\alpha\Omega$ dynamo is excited for $C_\alpha C_\Omega>2$ \citep{BS05}.
Because of $\nab\cdot\meanBB=0$, the resulting magnetic field is then
of the form $\meanBB(z)=(\meanB_x,\meanB_y,0)$, i.e., $\meanB_z=0$,
so it is not possible for the MRI to be excited.

\begin{figure}\begin{center}
\includegraphics[width=.7\columnwidth]{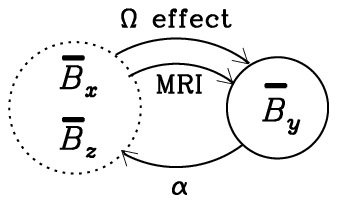}
\end{center}\caption{
Sketch illustrating the generation of $\meanB_y$ from $\meanB_x$
through the $\Omega$ effect and from $\meanB_z$ through the MRI,
and the generation of both $\meanB_x$ and $\meanB_z$ from $\meanB_y$
through the $\alpha$ effect.
}\label{psketch}\end{figure}

This would change if the dynamo also had an $x$ extent.
To see this, we consider for a moment a one-dimensional domain with
$\nab=(\partial_x,0,0)$.
In that case, an $\alpha^2$ dynamo with
$\meanBB(x)=(0,\meanB_y,\meanB_z)$ can be excited,
allowing $\meanB_z\neq0$.
It would be excited when $\alpha_0/\etaT k_{1x}\equiv C_\alpha k_1/k_{1x}>1$,
i.e., $C_\alpha>k_{1x}/k_1=L_z/L_x$.
\Fig{psketch} gives a graphical illustration of the generation of
$\meanB_y$ from $\meanB_x$ through the $\Omega$ effect and from $\meanB_z$
through the MRI, and the generation of both $\meanB_x$ and $\meanB_z$
from $\meanB_y$ through the $\alpha$ effect.

To allow for the possibility that in our two-dimensional domain
such a dynamo is preferred over one with $z$ extent, we choose
our domain to be oblate, e.g., $L_x/L_z=2$.
We solve the equations with the {\sc Pencil Code} \citep{PC} using
numerical resolutions between $64\times128$ and $256\times512$ mesh points,
i.e., the mesh spacings in the $x$- and $z$-directions are kept the same.

\subsection{Dynamo types in the R\"adler diagram}

\begin{figure}\begin{center}
\includegraphics[width=\columnwidth]{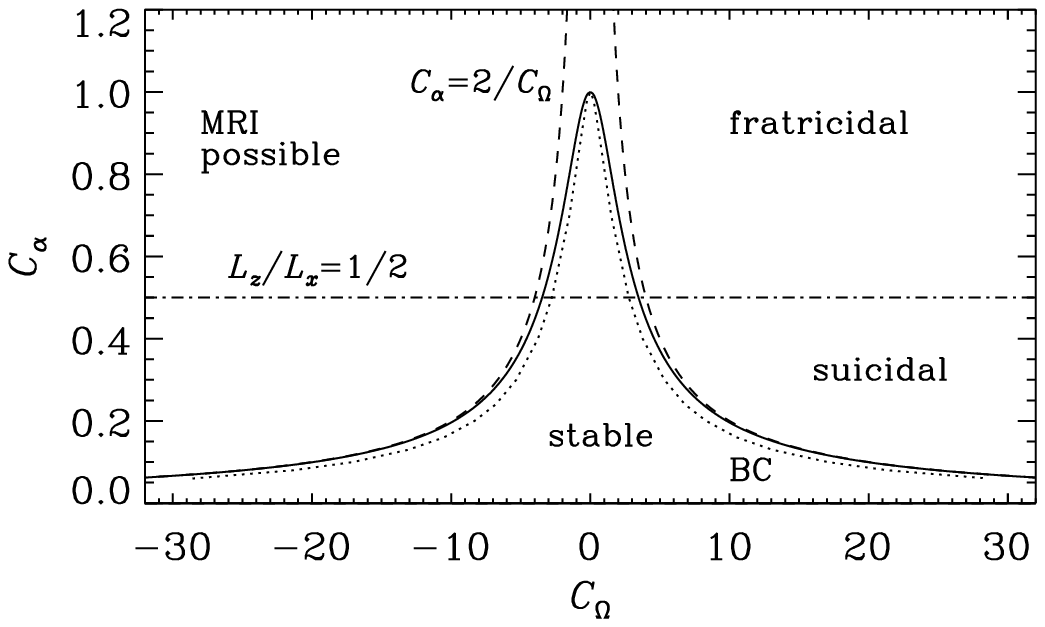}
\end{center}\caption{
R\"adler diagram for the $\alpha^2\Omega$ dynamo with $z$ extent
(solid line) and the $\alpha^2$ dynamo with $x$ extent in a domain
with $L_z/L_x=1/2$ (horizontal dash-dotted line).
The onset location in the pure $\alpha\Omega$ approximation
($C_\alpha C_\Omega=2$) is shown as dashed lines.
The case with the vertical field boundary condition is shown
as the dotted line and is marked BC.
}\label{pcrit}\end{figure}

It is convenient to discuss solutions in the $C_\alpha$--$C_\Omega$ plane;
see \Fig{pcrit}.
Such diagrams were extensively exploited by \cite{Raedler86}, which
is why we refer to such plots in the following as R\"adler diagrams.
R\"adler considered dynamos in spherical geometry where $\alpha$
changed sign about the equator, so the solutions were either symmetric
or antisymmetric about the equator.
In addition, they could be axisymmetric or nonaxisymmetric, and they
could also be oscillatory or stationary.

For a one-dimensional $\alpha^2\Omega$ dynamo with
periodic boundary conditions, the complex growth rate is
$(\alpha^2k^2-\ii k\alpha S)^{1/2}-\etaT k^2$ \citep{BS05}.
For the marginally excited state, we require the real part of the
complex growth rate to vanish.
This yields
\begin{equation}
C_\Omega=C_\alpha\sqrt{(2/C_\alpha^2-1)^2-1},
\end{equation}
which is the solid line shown in \Fig{pcrit}.
For the vertical field boundary condition, the dynamo is
slightly easier to excite; see the dotted line in \Fig{pcrit},
which has been obtained numerically.

The R\"adler diagram gives a graphical overview of the differences
between dynamos with positive and negative shear, i.e., positive and
negative values of $C_\Omega$.
The MRI is only possible for $C_\Omega<0$ (negative shear), while for
$C_\Omega>0$, we just expect ordinary $\alpha\Omega$ dynamo waves.
This expectation, however, does not apply to dynamos in periodic domains
with $\alpha_0=\const$, as was first found in the fully three-dimensional
turbulence simulations of \cite{Hubbard+11}.
Their $\alpha\Omega$ dynamo started off as expected, but at some point
during the early, nonlinear saturation phase of $\EEM^{X}$, the dynamo
wave stopped and a new solution emerged that had a cross-stream variation,
i.e., $\EEM^{X}$ became strong and suppressed $\EEM^{Z}$.

A similar type of exchange of dynamo solutions in the nonlinear regime was
first found by \cite{Fuchs+99} while investigating hydromagnetic dynamos
with Malkus--Proctor feedback in a sphere.
They found self-killing and self-creating dynamos due to the presence
of different stable flow patterns where the magnetic field causes the
solution to respond to a newly emerged flow pattern after the initial
saturation.
This was thus the first example of what then became known as a suicidal
dynamo.

\begin{figure}\begin{center}
\includegraphics[width=\columnwidth]{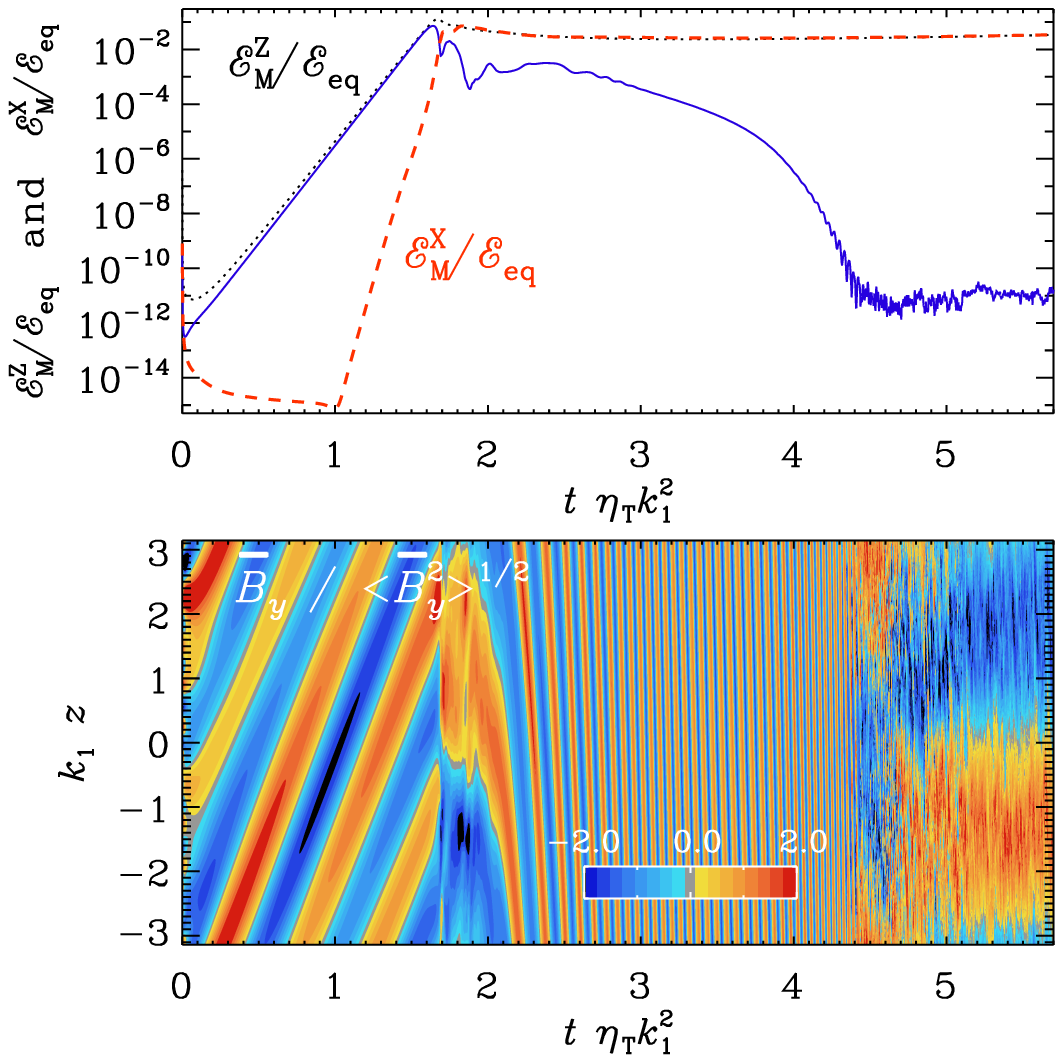}
\end{center}\caption{
Time dependence of $\EEM$ (dotted black line), $\EEMZ$ (solid blue line),
and $\EEMX$ (dashed red line), all normalized by $\EEMeq$, and $\meanB_y$
versus $t$ and $z$ for a fratricidal dynamo (Run~F) with $C_\alpha=1$,
$C_\Omega=150$, $q=-3/2$ (positive shear), and $\Beq\to\infty$ (no $\alpha$
quenching).
Here, $\meanB_y$ has been normalized by its instantaneous rms values so as
to see the dynamo wave also during the early exponential growth phase and
during the late decay phase.
}\label{pfratri_bm}\end{figure}

In analogy with the suicidal dynamos, the dynamos found by \cite{Hubbard+11}
were called fratricidal dynamos.
This property of dynamos in a periodic domain emerged as a problem because
$\alpha\Omega$ dynamos in a periodic domain could only be studied for
a limited time interval before they disappeared \citep{KB16}.

\section{Results}

We begin with a discussion of fratricidal and suicidal dynamos, but
emphasize that these have so far only been found in periodic systems
for $C_\Omega>0$, i.e., for positive shear.
Thus, to examine the effect of the MRI, we compare solutions with
positive and negative values of $C_\Omega$ using both periodic and
nonperiodic domains.

\begin{figure}\begin{center}
\includegraphics[width=\columnwidth]{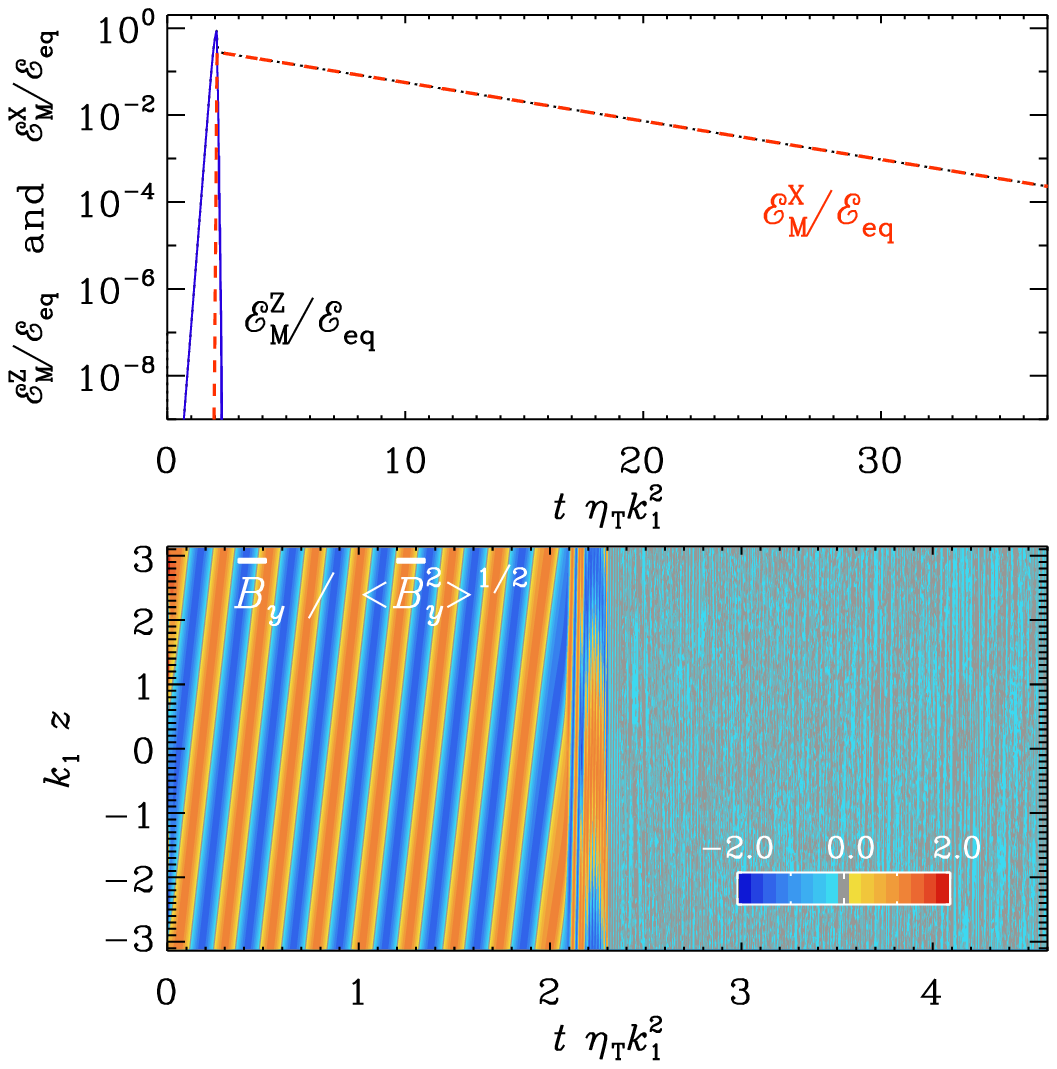}
\end{center}\caption{
Similar to \Fig{pfratri_bm}, but for a suicidal dynamo with
$C_\alpha=0.49$ and $C_\Omega=7.5$ (Run~B).
}\label{psuicidal_bm}\end{figure}

\begin{figure*}\begin{center}
\includegraphics[width=.95\textwidth]{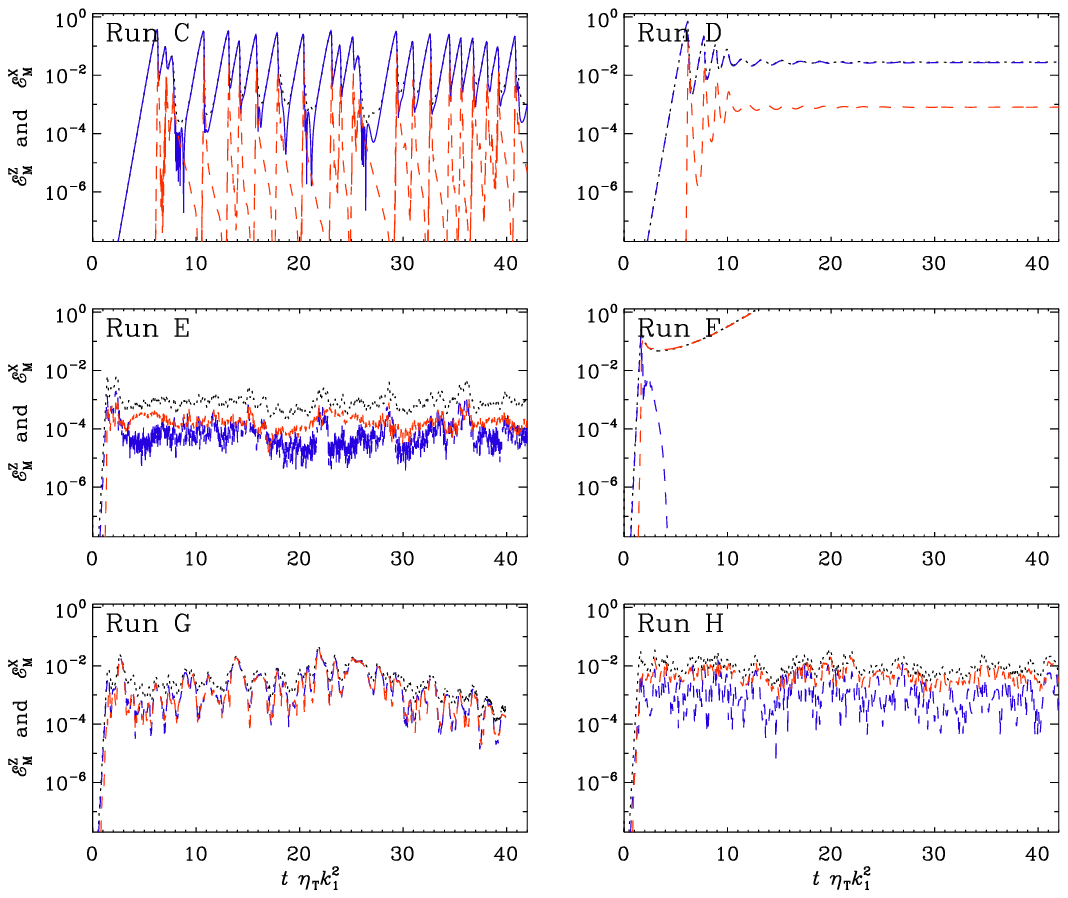}
\end{center}\caption{
Comparison of solutions for $C_\Omega<0$ (Runs~C, E, and G; left panels)
and $C_\Omega>0$ (Runs~D, F, and H; right panels) for periodic boundary
conditions (top and middle) and vertical field boundary conditions (bottom).
As in the upper panels of \Figs{pfratri_bm}{psuicidal_bm}, $\EEM$ (dotted
black line), $\EEMZ$ (solid blue line), and $\EEMX$ (dashed red line),
normalized by $\EEMeq$, are shown versus $t$.
}\label{pcomp_bm}\end{figure*}

\subsection{Fratricidal and suicidal mean-field dynamos}

Here, we show that both fratricidal and suicidal dynamos can also occur
in a mean-field context; see \Figs{pfratri_bm}{psuicidal_bm}.
The $\alpha^2$ sibling is here possible because $C_\alpha>L_z/L_x=0.5$.
This is shown in \Fig{pfratri_bm}, where we plot $\EEMZ$ and $\EEMX$
versus time, and $\meanB_y$ versus $t$ and $z$.
In the following, this case is referred to as Run~F.
We see that $\EEMZ$ grows exponentially, starting from a weak seed
magnetic field.
The $zt$ diagram in \Fig{pfratri_bm} shows the usual dynamo waves.
When the dynamo approaches saturation, $\EEMX$ also begins to grow
exponentially, but at a rate that it is much larger than the growth rate
of $\EEMZ$.
When $\EEMX$ reaches about $10^{-3}\EEMeq$, $\EEMZ$ 
declines rapidly and is then overtaken by $\EEMX$.
At that moment, the dynamo waves cease and a new transient commences
with a rapidly varying time dependence, but at a very low amplitude;
see the $zt$ diagram of \Fig{pfratri_bm} for $2.5<t\etaT k_1^2<4.5$.

For $C_\alpha<0.5$, the $\alpha^2$ sibling with $\EEM^{X}\neq0$ is impossible.
Surprisingly, it turned out that the $\alpha\Omega$ dynamo can then
still be killed by a secondary $\EEM^{X}$, but such a state with
$\EEM^{X}\neq0$ cannot be sustained and decays on an ohmic time scale;
see \Fig{psuicidal_bm} for Run~B.
It is therefore an example of a suicidal dynamo.
We see that $\EEMX$ decays toward zero, and that the dynamo wave then
just disappears.
By that time, $\EEMZ$ has already become very small and has
disappeared within the noise.

\subsection{Comparison of positive and negative shear}
\label{PositiveNegativeShear}

In our fully nonlinear simulations, it is difficult to say whether the
MRI was important and had any particular effect.
Unlike an actual physical system, a simulation allows us in principle
to identify the effect of each term in the equations by modifying it
artificially.
Shear is of course crucial for the MRI, but shear is also crucial for
the $\Omega$ effect.
In our Cartesian system, however, dynamos with prescribed shear and no
other dynamics from the momentum equation have analogous properties,
i.e., the same rms values of the magnetic field components and relative
phase shifts between them, which transform in a known way
\citep{Yoshimura75, Yoshimura76, Stix76}.
Thus, to identify the effect of the MRI, it is convenient to
compare solutions for positive and negative shear.
In \Fig{pcomp_bm}, we plot the time evolutions of $\EEM$, $\EEM^{X}$,
and $\EEM^{Z}$ for Runs~C--G with different values of $C_\alpha$ and
$C_\Omega$, as well as periodic and vertical field boundary conditions.
We see that, regardless of the boundary conditions, the cases with
negative shear, where the MRI is possible, tend to have less magnetic energy
than the cases with positive shear.

Various parameters related to the flow of energy are summarized
in \Tab{Tenergetics}.
We see that $W_\mathrm{L}$ is positive, i.e., magnetic energy
goes into kinetic energy.
This is typical of the MRI and has been found previously for
turbulent MRI dynamos \citep{BNST95}.
But we also see that whenever $C_\Omega$ is negative and the MRI is
excited, $W_\mathrm{M}$, $W_\mathrm{K}$, $\epsM$, $\epsK$, and in most
cases also $W_\mathrm{L}$ are much larger than for
positive values of $C_\Omega$, when the MRI does not operate.
In the latter case, when only the standard $\Omega$
effect operates, $W_\mathrm{K}$ is often even negative.
Note also that $W_\mathrm{P}$ is not being given, because its value
is very small.
Likewise, $\dot{\mathcal{E}}_\mathrm{M}$ and $\dot{\mathcal{E}}_\mathrm{K}$
are small and not listed, but are still included in the calculation of
the total
\begin{equation}
\mbox{gain}=W_\mathrm{M}+W_\mathrm{K}+W_\alpha+W_\mathrm{P}
\end{equation}
and
\begin{equation}
\mbox{loss}=\epsM+\epsK+\dot{\mathcal{E}}_\mathrm{M}
+\dot{\mathcal{E}}_\mathrm{K}.
\end{equation}
Both the total gain and the total loss balance each other nearly
perfectly.
In \Tab{Tenergetics}, we also give the nondimensional growth rate,
$\tilde{\lambda}\equiv\lambda/\etaT k_1^2$, where
$\lambda=\dd\ln\meanBrms/\dd t$ is the physical growth rate.

\begin{table*}\caption{
Summary of Runs~A--{\AA} for $q=3/2$ and Runs~c--j for $q=1$.
The BC column gives 0 (1) for periodic (vertical field) boundary conditions.
For runs without $\alpha$ quenching, we have $\mathcal{B}_\mathrm{eq}^{-1}=0$.
$\EEM$ and $\EEK$ are given in units of $\rho_0 S^2/k_1^2$.
The energy fluxes $W_\mathrm{M}$, $W_\mathrm{K}$, $W_\alpha$, $W_\mathrm{L}$, $\epsM$, $\epsK$,
as well as gain and losses are in units of $\etaT k_1^2\EEM$.
The last column denotes the nondimensional growth rate $\tilde{\lambda}\equiv\lambda/\etaT k_1^2$.
}\hspace{-20mm}\vspace{12pt}\centerline{\begin{tabular}{cccc rccr rrrr rrrr rrr}
Run& BC & $\mathcal{B}_\mathrm{eq}^{-1}$ & $C_\alpha$ & $C_\Omega$ & $\EEM$ & $\EEK$ & $W_\mathrm{M}$ &
	$W_\mathrm{K}$ & $W_\alpha$ & $W_\mathrm{L}$ & $\epsM$ & $\epsK$ & gain  &  loss & $\tilde{\lambda}\;$ \\
\hline
A & 0 &  0 & 0.49 &$  -7.5 $&  2.87 &  0.37 &  2.5 &$  0.290 $&  0.500 &  0.16 &  2.8 &  0.59 &  3.3 &  3.3 &  0.4 \\
B & 0 &  0 & 0.49 &$   7.5 $&  1.00 &  1.93 &  0.0 &$  0.000 $&  0.490 &  0.01 &  0.5 &  0.01 &  0.5 &  0.5 &  0.4 \\
C & 0 &  0 & 0.20 &$   -15 $&  1.99 &  0.08 &  2.4 &$  0.050 $&  0.085 &  0.06 &  2.5 &  0.23 &  2.6 &  2.6 &  0.2 \\
D & 0 &  0 & 0.20 &$    15 $&  0.62 &  0.02 &  2.0 &$ -0.001 $&  0.080 &  0.04 &  2.0 &  0.04 &  2.0 &  2.1 &  0.2 \\
E & 0 &  0 & 1.00 &$  -150 $&  0.04 &  0.27 & 37.0 &$ 10.000 $&  2.000 &  6.70 & 33.0 & 19.00 & 50.0 & 50.0 &  7.5 \\
F & 0 &  0 & 1.00 &$   150 $&  1.26 &  0.68 &  0.3 &$  0.007 $&  1.700 &  0.37 &  1.8 &  0.35 &  2.0 &  2.2 &  7.9 \\
G & 1 &  0 & 1.00 &$  -150 $&  0.18 &  0.28 & 25.0 &$  6.900 $&  1.300 &  2.50 & 25.0 &  8.90 & 33.0 & 33.0 &  7.0 \\
H & 1 &  0 & 1.00 &$   150 $&  0.24 &  0.22 &  8.8 &$ -0.310 $&  0.780 &  3.50 &  6.6 &  3.50 &  9.2 &  9.7 &  7.2 \\
I & 1 &  0 & 0.20 &$  -150 $&  0.12 &  0.24 &  8.1 &$  0.600 $&  0.043 &  2.30 &  7.2 &  3.50 &  8.8 & 10.0 &  2.0 \\
J & 1 &  0 & 0.20 &$   150 $&  0.33 &  0.12 &  3.6 &$ -0.012 $&  0.030 &  0.73 &  3.3 &  0.89 &  3.6 &  4.1 &  1.8 \\
\hline
K & 0 &  1 & 0.49 &$  -7.5 $&  0.15 &  0.00 &  1.8 &$  0.000 $&  0.170 &  0.00 &  2.0 &  0.00 &  2.0 &  2.0 &  0.4 \\
L & 0 &  1 & 0.49 &$   -30 $&  1.07 &  0.00 &  2.0 &$ -0.000 $&  0.028 &  0.00 &  2.0 &  0.00 &  2.0 &  2.0 &  1.7 \\
M & 0 &  1 & 0.49 &$   -75 $&  0.04 &  0.22 & 10.0 &$  0.850 $&  0.250 &  2.80 &  7.9 &  4.30 & 11.0 & 12.0 &  3.0 \\
N & 0 &  1 & 0.49 &$  -150 $&  0.04 &  0.17 & 18.0 &$  1.500 $&  0.280 &  3.00 & 15.0 &  6.10 & 19.0 & 19.0 &  4.6 \\
O & 0 &  1 & 0.49 &$  -300 $&  0.02 &  0.23 & 31.0 &$  3.100 $&  0.330 &  7.80 & 24.0 & 14.00 & 34.0 & 34.0 &  7.4 \\
\hline
P & 0 & 10 & 0.49 &$   -30 $&  0.01 &  0.00 &  2.0 &$ -0.000 $&  0.027 &  0.00 &  2.0 &  0.00 &  2.0 &  2.0 &  1.6 \\
Q & 0 & 10 & 0.49 &$   -75 $&  0.03 &  0.00 &  2.0 &$ -0.000 $&  0.008 &  0.00 &  2.0 &  0.00 &  2.0 &  2.0 &  3.0 \\
R & 0 & 10 & 0.49 &$  -300 $&  0.12 &  0.00 &  2.1 &$ -0.000 $&  0.001 &  0.00 &  2.1 &  0.00 &  2.1 &  2.1 &  7.4 \\
S & 0 & 10 & 0.49 &$  -750 $&  0.10 &  0.00 &  4.0 &$  0.000 $&  0.010 &  0.04 &  4.0 &  0.04 &  4.1 &  4.0 & 13.4 \\
T & 0 & 10 & 0.49 &$ -1500 $&  0.09 &  0.01 &  7.1 &$  0.008 $&  0.003 &  0.21 &  6.9 &  0.23 &  7.1 &  7.0 & 20.7 \\
U & 0 & 10 & 0.49 &$ -3000 $&  0.04 &  0.00 & 16.0 &$  0.880 $&  0.021 &$-0.07$& 16.0 &  0.86 & 17.0 & 19.0 & 30.9 \\
\hline
V & 0 &100 & 0.49 &$  -300 $&  0.00 &  0.00 &  2.1 &$ -0.000 $&  0.002 &  0.00 &  2.1 &  0.00 &  2.1 &  2.1 &  6.8 \\
W & 0 &100 & 0.49 &$  -750 $&  0.00 &  0.00 &  2.1 &$ -0.000 $&  0.000 &  0.00 &  2.1 &  0.00 &  2.1 &  2.1 & 11.6 \\
X & 0 &100 & 0.49 &$ -1500 $&  0.01 &  0.00 &  2.1 &$ -0.000 $&  0.000 &  0.00 &  2.1 &  0.00 &  2.1 &  2.1 & 16.3 \\
Y & 0 &100 & 0.49 &$ -3000 $&  0.01 &  0.00 &  2.6 &$ -0.000 $&  0.000 &  0.00 &  2.3 &  0.00 &  2.6 &  2.6 & 20.6 \\
Z & 0 &100 & 0.49 &$ -7500 $&  0.01 &  0.00 &  4.0 &$  0.000 $&  0.000 &  0.01 &  3.2 &  0.01 &  4.0 &  4.0 & 20.8 \\
\AA&0 &100 & 0.49 &$-15000 $&  0.01 &  0.00 &  8.5 &$  0.000 $&  0.000 &  0.02 &  6.6 &  0.01 &  8.5 &  8.5 & 19.1 \\
\hline
c & 0 &  0 & 0.20 &$   -15 $&  0.71 &  7.02 &  2.4 &$ -0.120 $&  0.084 &  0.03 &  2.5 &  1.60 &  2.4 &  3.8 &  0.2 \\
d & 0 &  0 & 0.20 &$    15 $&  0.81 &  0.01 &  1.9 &$  0.000 $&  0.082 &  0.04 &  2.0 &  0.04 &  2.0 &  2.0 &  0.2 \\
e & 0 &  0 & 1.00 &$  -150 $&  0.29 &  1.79 & 25.0 &$  6.500 $&  1.900 &  8.50 & 17.0 & 15.00 & 33.0 & 32.0 &  7.9 \\
f & 0 &  0 & 1.00 &$   150 $&  0.89 &  0.97 &  0.3 &$ -0.008 $&  1.800 &  0.64 &  1.9 &  0.33 &  2.0 &  2.5 &  7.9 \\
i & 1 &  0 & 0.20 &$  -150 $&  0.57 &  0.89 &  6.5 &$  0.520 $&  0.029 &  1.90 &  5.7 &  2.30 &  7.0 &  7.6 &  2.0 \\
j & 1 &  0 & 0.20 &$   150 $&  0.38 &  0.14 &  3.7 &$ -0.011 $&  0.030 &  0.85 &  3.3 &  0.95 &  3.8 &  4.1 &  1.8 \\
\label{Tenergetics}\end{tabular}}\end{table*}

\begin{figure*}\begin{center}
\includegraphics[width=\textwidth]{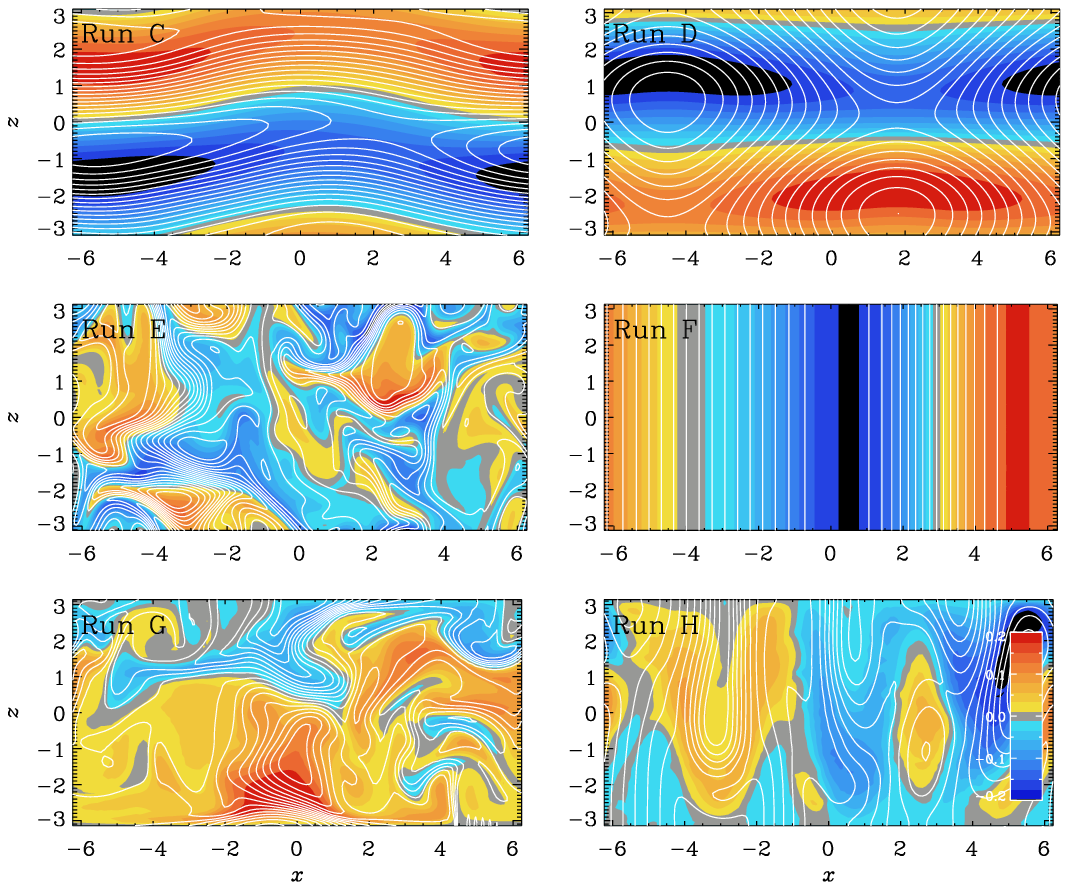}
\end{center}\caption{
Visualizations of field lines of $(\meanB_x,\meanB_z)$ in the $x$--$z$
plane on top of a color-scale representation of $\meanB_y$ for Runs~C--H,
where blue (red) shades refer to negative (positive) values.
}\label{ppvar_comp}\end{figure*}

Interestingly, the ratio $\epsK/\epsM$, which is known to scale with the
microphysical magnetic Prandtl number in direct numerical simulations of
forced turbulence \citep{Bra14}, varies widely in the present mean-field
calculations.
It is less than unity, and often much less than unity.
On the other hand, not much is known about the scaling of this dissipation
ratio for MRI-driven turbulence.
In the old simulations of \cite{BNST95}, this ratio was found to be even
slightly larger than unity.
Given that we present only a coarse coverage of a fairly large parameter
space in the R\"adler diagram, it is possible that there are some
relationships that cannot presently be discerned.

\begin{figure}\begin{center}
\includegraphics[width=\columnwidth]{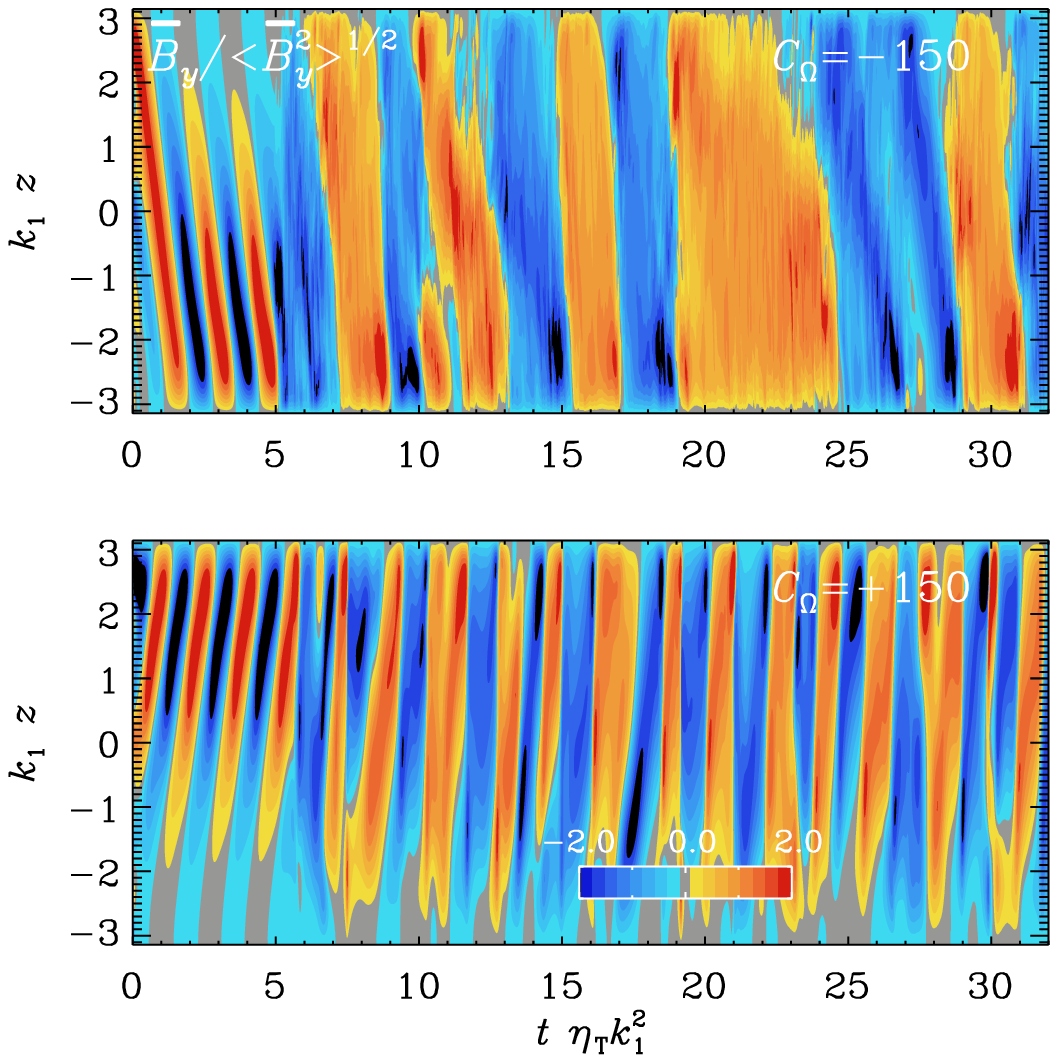}
\end{center}\caption{
Mean magnetic field evolution in a $zt$ diagram for simulations with
vertical field boundary conditions in the $z$-direction for Runs~I and
J with $C_\Omega=-150$ (upper panel) and $C_\Omega=+150$ (lower panel),
respectively, using $C_\alpha=0.2$.
}\label{pboundaries}\end{figure}

\subsection{Magnetic field structures}

It is instructive to inspect the magnetic field structures of individual
snapshots.
This is shown in \Fig{ppvar_comp}, where we present visualizations of field
lines in the $x$--$z$ plane together with a color-scale representation of $\meanB_y$ for Runs~C--H.
In our two-dimensional case, field lines are shown as contours of $\meanA_y$.
Runs~C and D have a predominantly vertical dependence, which was already
indicated by the dominance of $\EEMZ$ over $\EEMX$ in \Fig{pcomp_bm}.
As we have seen before, the MRI operates in Run~C, causing
some residual $x$ dependence in the field, as manifested by the wavy
pattern.

Run~F is the complete opposite of Run~D, because now there is only a
pure $x$ dependence.
Again, this was also already indicated in \Fig{pcomp_bm} through the
dominance of $\EEMX$ over $\EEMZ$.
This dramatic difference is explained by the value of $C_\alpha=1$,
which is now large enough for an $\alpha^2$ dynamo with an $x$ extent to
be excited.

For negative shear, on the other hand, Runs~C and E also show a change
from a predominantly $z$-dependent field for small values of $C_\alpha$
(Run~C) to a predominantly $x$-dependent field for large values of
$C_\alpha$ (Run~E).
However, unlike the fratricidal dynamo for positive shear, where
$\EEMZ$ is completely killed, it is here only partially suppressed;
see \Fig{pcomp_bm}.
We could therefore call such a dynamo a narcissistic one.
The dominant $x$ dependence of the magnetic field is also evident
from \Fig{ppvar_comp}.

Runs~E and G show predominantly small-scale structures.
There is no strong difference between the periodic and nonperiodic runs,
except that the field lines are now purely vertical on the boundaries.
It is these small-scale structures that are responsible for the enhanced
dissipation and ultimately for the decreased efficiency of the dynamo
process in the presence of the MRI.

Run~H also displays small-scale structures, but these are not related
to the MRI, which is absent in this run with positive shear.
Here, the existence of small-scale structures is probably related to
the presence of boundaries in the $z$-direction.
They lower the excitation conditions for dynamo action with magnetic
field dependence in the $z$-direction, but there could also be other
reasons for the existence of small-scale structures in this case.

In \Tab{Tenergetics}, we also give the results of runs (Runs~c--f, as well as i and j)
where $q=1$ instead of 3/2 but $S$ is unchanged, so $\Omega$ is then
chosen to be 3/2.
These runs are otherwise similar to Runs~C--F, as well as I and J, respectively.

\subsection{Simulations with vertical boundary conditions}

Next, we study the mean magnetic field evolution for simulations
with vertical field boundary conditions in the $z$-direction.
The resulting $zt$ diagrams are shown in \Fig{pboundaries}
for Runs~I and J with $C_\Omega=-150$ and $+150$, respectively, using $C_\alpha=1$.
Note that during the early kinematic phase there is clear evidence
for dynamo waves migrating in the negative (positive) $z$-direction for
negative (positive) values of $C_\Omega$.

Comparing Runs~F and G in \Tab{Tenergetics}, they have the same
parameters, but Run~G has vertical field boundary conditions.
We see that $W_\mathrm{K}$ is much larger in Run~G than in Run~F.
Also $W_\mathrm{L}$ is significantly larger in Run~G, but the
difference is here not quite as large.
This is presumably caused by the existence of small-scale structures in
Run~G, while Run~F has essentially only a one-dimensional field structure
at late times.

\subsection{Transition from $\Omega$ effect to MRI}

When $C_\Omega$ is small enough, the turbulent magnetic diffusivity may
be too large for the MRI to be excited, 
as the magnetic diffusion rate might exceed the typical growth rate of the instability,
which is of the order of $\Omega$.
This idea assumes that the magnetic field is held fixed, but this is not
true when the magnetic field is still being amplified by dynamo action
and saturation by the large-scale Lorentz force has not yet been achieved.
Therefore, since the magnetic field might still be growing, it would not
be surprising if the MRI occurred even for small values of $C_\Omega$,
corresponding to larger magnetic diffusion rates.

\begin{figure}\begin{center}
\includegraphics[width=\columnwidth]{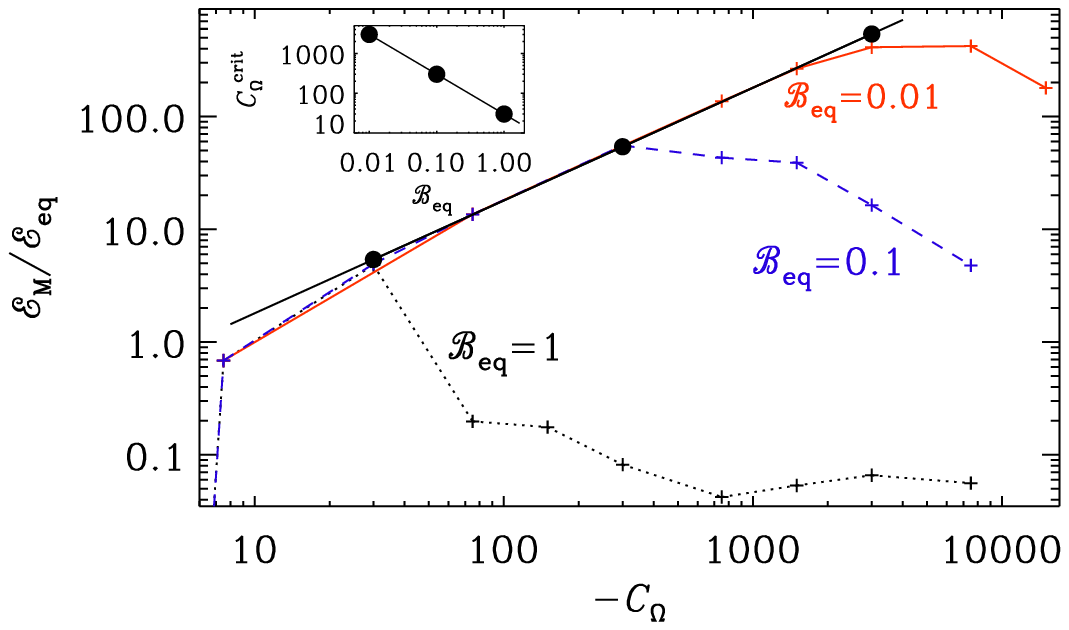}
\end{center}\caption{
Dependence of $\EEM/\EEMeq$ on $C_\Omega$ for $\mathcal{B}_\mathrm{eq}=1$
(black dotted line), $0.1$ (blue dashed line), and $0.01$ (red solid line)
using $C_\alpha=0.49$ in all cases.
The black solid line denotes $\EEM/\EEMeq=0.18\,|C_\Omega|$ and the
filled circles on this line denote the approximate values where $\EEM$
departs from the linearly increasing trend with $|C_\Omega|$.
The inset shows the dependence of $C_\Omega^\mathrm{crit}$ on
$\mathcal{B}_\mathrm{eq}$.
}\label{pEEM_vs_Com}\end{figure}

To facilitate dynamo saturation at a lower magnetic field strength, and
therefore a regime with $C_\Omega<0$ without MRI, we now invoke $\alpha$
quenching with finite values of $\Beq$.
(The case without $\alpha$ quenching corresponds to $\Beq\to\infty$.)
We have performed numerical experiments for different values of $\Beq$
and $C_\Omega$.
It turns out that for a fixed value of $\Beq$, there is a critical value
of $C_\Omega$ above which the MRI commences.
This is shown in \Fig{pEEM_vs_Com}, where we plot the mean magnetic
energy density versus $-C_\Omega$ (for $C_\Omega<0$) and a fixed value
of $C_\alpha=0.49$.
We see that $\EEM$ increases approximately linearly with $|C_\Omega|$ and
has the same value when normalized by the respective value of $\EEMeq$.
Because the normalized values $\EEM/\EEMeq$ are the same for different
values of $|C_\Omega|$ and different values of $\EEM$, this saturation
dependence is a consequence of $\alpha$ quenching.
Above a certain value of $|C_\Omega|$, however, the increasing trend
stops and $\EEM$ begins to decline with increasing values of $|C_\Omega|$.
We associate this with the onset of the MRI.

The MRI onset occurs for smaller values of $|C_\Omega|$ when
$\mathcal{B}_\mathrm{eq}$ is large.
This is understandable, because for large values of
$\mathcal{B}_\mathrm{eq}$, $\alpha$ quenching commences
only for stronger magnetic fields.
Therefore, magnetic field saturation can be accomplished by the MRI
before $\alpha$ quenching would be able to act.
From the inset of \Fig{pEEM_vs_Com}, we find quantitatively
\begin{equation}
C_\Omega^\mathrm{crit}\approx30\,\mathcal{B}_\mathrm{eq}^{-1}.
\end{equation}
Thus, although $C_\Omega<0$, the standard $\Omega$ effect is expected
to operate in the range
\begin{equation}
2/C_\alpha\la|C_\Omega|\la C_\Omega^\mathrm{crit},
\end{equation}
and the MRI is only possible for values of $|C_\Omega|$ larger than
$C_\Omega^\mathrm{crit}$.

\subsection{Comparison with earlier work}
\label{EarlierWork}

Let us now discuss whether the MRI might have been excited in previously
published work.
Hydromagnetic models with $\alpha$ and $\Lambda$ effects were considered
by \cite{Bra+92} using spherical geometry.
The sign of $C_\Omega$ was determined by the sign of the $\Lambda$ effect.
Their $C_\Omega$ is defined based on the stellar radius $R$ and can
therefore not directly be compared with the $C_\Omega$ used in the
present work.
Also, given that the differential rotation emerges as a result of the
$\Lambda$ effect and is already affected by the magnetic field, their
$C_\Omega$ is an output parameter.

In their Run~T5 of model A$-$, they found $C_\Omega=-474$, while for
their Run~T7 of model~A+, they found $C_\Omega=+939...+1010$.
The magnetic field in this model was oscillatory, which explains the
existence of a range of $C_\Omega$.

To address the question whether the MRI was operational in their model A$-$,
we use the fact that the resulting kinetic energy is large for
negative shear, compared to the case of positive
shear; see \Sec{PositiveNegativeShear}.
\cite{Bra+92} specified the decadic logarithms and found a kinetic energy of
$\EEK=10^{2.40}$ for their model A$-$ and $\EEK=10^{1.65...1.74}$ for
their model A+.
This suggests that the MRI was operational in their model~A$-$.
Note also that $|C_\Omega|$ was smaller in their Run~T5 compared to Run~T7.
If the driving of kinetic energy were related only to the magnitude of the shear,
one would expect the opposite trend.
This confirms that the increased kinetic energy in model~A$-$ was indeed
due to the MRI.

To further assess the excitation of the MRI, we can also estimate their
effective value of $\vA k_1/\Omega$.
Using $\vA\approx\sqrt{2\EEM/\rho_0}\approx150$, $k_1=2\pi/0.3R\approx20$,
$\Omega=\Ta^{1/2}\etaT/2R^2\approx2700$, where $\Ta=3\times10^7$ is the
turbulent Taylor number, and $\Pm=1$, we find $\vA k_1/\Omega\approx1$,
which is consistent with the MRI being excited.

\begin{figure}\begin{center}
\includegraphics[width=\columnwidth]{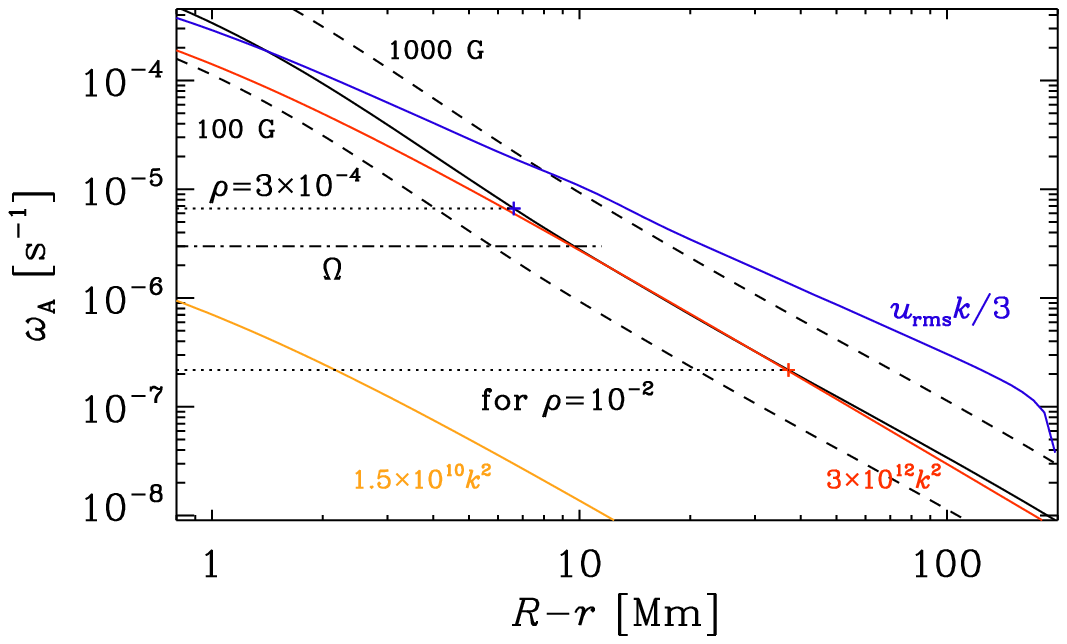}
\end{center}\caption{
Depth dependence of the Alfv\'en frequency
for $\meanBrms=300\G$ (solid black line) using the mixing length model
of \cite{Spruit74}.
Also shown are the values for $\meanBrms=1000\G$ and $\meanBrms=100\G$
(upper and lower dashed lines), as well as $\urms k/3$ (blue),
$3\times10^{12}k^2\cm\s^{-1}$ (red line),
and $1.5\times10^{10}k^2\cm\s^{-1}$ (orange line).
The horizontal dotted lines correspond to the depths
of $7\Mm$ (where $\rho=3\times10^{-4}\s^{-1}$) and
$40\Mm$ (where $\rho=10^{-2}\s^{-1}$), and
$\Omega=3\times10^{-6}\s^{-1}$ marks the solar angular velocity.
}\label{prho}\end{figure}

\subsection{Estimates for the Sun}

Let us now estimate some relevant parameters for the Sun.
A similar comparison was already presented by \cite{Vasil+24}.
For the MRI to be excited, the Alfv\'en frequency,
$\omega_\mathrm{A}=\vA k$, must not exceed the rotational shear frequency,
$\sqrt{2q}\,\Omega$, but it must also be larger than the turbulent
diffusion rate, $\etaT k^2$, so
\begin{equation}
\etaT k^2<\omega_\mathrm{A}<\sqrt{2q}\,\Omega.
\end{equation}
For the solar NSSL, we have $q=-\partial\ln\Omega/\partial\ln\varpi=1$ \citep{Barekat+14}.
For $k$, we estimate $k\approx1/\ell$, where $\ell$ is the local mixing length,
which is also approximately equal to the depth, $R-r$, where $R$ is the
solar radius and $r$ is the local radius; see \Tab{Tnondim} for a summary
of some nondimensional parameters.

\begin{table}\caption{
Summary of solar estimates for some nondimensional parameters.
}\hspace{-20mm}\vspace{12pt}\centerline{\begin{tabular}{cccc rccr rrrr rrrr rrr}
depth & $\mathcal{B}_\mathrm{eq}^{-1}$ & $C_\Omega$ & $\EEM/\rho_0 S^2$ \\
\hline
$ 7\Mm$ & 0.3 & 0.3 & 6     \\
$37\Mm$ & 9   & 9   & 0.006 \\
\hline
\label{Tnondim}\end{tabular}}\end{table}

In \Fig{prho}, we plot the dependence of $\omega_\mathrm{A}$
on the depth $R-r$, where the radial dependence of $\ell$ and $\rho$ has been
obtained from the solar mixing length model of \cite{Spruit74}.
Here, we also present estimates of $\etaT k^2$,
where we assume either a constant $\etaT$
($3\times10^{12}\cm^2\s^{-1}$) or we use the result of \cite{Sur+08}:
\begin{equation}
\etaT=\urms/3k.
\label{etaT-estimate}
\end{equation}
Both estimates show a similar dependence on depth.
The value $\etaT=3\times10^{12}\cm^2\s^{-1}$ is motivated by a similar
one for the turbulent heat diffusivity; see \cite{Krivodubskii84}.
\cite{Vasil+24} also adopted turbulent viscosities and magnetic
diffusivities, which they specified as
$10^{-6}\,R^2\Omega_0\approx1.5\times10^{10}\cm^2\s^{-1}$.
This value, which is 200 times smaller than our estimates above,
corresponds to the orange line in \Fig{prho}.

Using for the mean field of the Sun $\meanBrms=300\G$ \citep{Bra05ApJ, Vasil+24},
we have $\vA=50\m\s^{-1}$ and $\oA=7\times10^{-6}\s^{-1}$ at a
depth of $7\Mm$ where $\rho\approx3\times10^{-4}\g\cm^{-3}$, and
$\vA=8\m\s^{-1}$ and $\oA=2\times10^{-7}\s^{-1}$ at a depth of $40\Mm$
where $\rho\approx10^{-2}\g\cm^{-3}$.
These values bracket the value of $\Omega$, so the MRI might
be viable somewhere in this range.
However, different estimates for the turbulent diffusion rate $\urms k/3$
(shown in blue) and $3\times10^{12}\cm^2\s^{-1}\,k^2$ (shown in red)
lie tightly at $\oA$ or even exceed it
at nearly all depths, making the MRI difficult to excite.
Furthermore, if we estimated $k=2\pi/\ell$ instead of just $1/\ell$, 
$\oA$ would attain much higher values and
the MRI would not be excited.

Before concluding, let us comment on solar estimates for some of the
nondimensional parameters defined in \Sec{Input+OutputParameters}.
We see from \Eq{calBdef} that $\oA/\Omega\equiv\mathcal{B}_\mathrm{eq}$.
\Fig{prho} shows that this ratio varies between 3 for $R-r=6\Mm$
and 0.1 for $R-r=40\Mm$; see also \Tab{Tnondim}, where we give
$\mathcal{B}_\mathrm{eq}^{-1}$ (as in \Tab{Tenergetics})
along with other nondimensional parameters.
Note also that, in the parameter range of interest, we have
$\mathcal{B}_\mathrm{eq}^{-1}\approx C_\Omega$.
This is because for $\Brms=300\G$, we have $\etaT k^2\approx\oA$, as
was already seen from the agreement between the red and black lines
in \Fig{prho}.
The nondimensional magnetic energies, $\EEM/\rho_0 S^2$, are given by
$\mathcal{B}_\mathrm{eq}^2/2q^2$ and lie in the range between 6 (closer
to the surface) and 0.006 near the  bottom of the NSSL.

\section{Conclusions}

The MRI can only work with negative shear, i.e., when $C_\Omega<0$.
We find that for our models without $\alpha$ quenching and with
$C_\Omega<0$, i.e., when the MRI can operate, the kinetic energy
production ($W_\mathrm{K}$) and dissipation ($\epsK$) are large
compared to the case where $C_\Omega>0$.
As discussed in \Sec{EarlierWork},
the models of \cite{Bra+92}, where no $\alpha$ quenching was included,
do show enhanced kinetic energy production for negative shear.
This suggests that the MRI might indeed have been operating in those models.

Our work has also shown that the MRI can work even for small shear
parameters when the magnetic field strength is limited by the
large-scale Lorentz force only.
However, mechanisms such as $\alpha$ quenching related to the backreaction
of the Lorentz force from the small-scale field can prevent the MRI from
occurring for small shear parameters.
This $\alpha$ quenching limits the magnetic field strength to values
below the critical one where
the magnetic diffusion rate exceeds the growth rate of the MRI.

Finally, we discussed whether or not the MRI could play
a role in the Sun.
Our estimates suggest that turbulent diffusion is likely too large for
the MRI to be excited, but the estimates are uncertain because they
depend on the magnetic field strength and the value of the wavenumber.
If we assumed it were $2\pi/\ell$, the MRI would definitely be
ruled out, while for $k=1/\ell$, it would be right at the limit for
$\meanBrms=300\G$.
This value of the magnetic field strength is also what was considered
by \cite{Bra05ApJ}, and it is compatible with what was assumed by
\cite{Vasil+24}, who discussed values in the range between $100$
and $1000\G$.

It will be of interest to confront our specific findings based on
two-dimensional mean-field calculations with those based on direct
three-dimensional simulations.
So far, the only paper that discusses such a comparison for the MRI is
that of \cite{Vaisala+14}, who demonstrated that the onset of the MRI
is consistent with what is expected from mean-field estimates.
In those cases, there was an imposed magnetic field, which would now
need to be replaced by a dynamo-generated one.
A problem in reproducing mean-field results in direct numerical
simulations is the strength and coherence of the large-scale magnetic
field; see \cite{Hughes18} for a critical review.
A particularly crucial point is the possibility of what has been referred
to as catastrophic quenching; see \cite{Bra18} for another review, and
\cite{Kapy25} for a more recent one.
A leading candidate for alleviating such quenching is magnetic helicity
fluxes, which have been computed in numerous studies since the work
of \cite{VC01}.
Those suggest that catastrophic dynamo quenching is alleviated by the 
shear-induced hemispheric small-scale magnetic helicity fluxes.
They have a strong effect, as was recently shown in \cite{BV25}.
It would therefore be of interest to repeat such studies in cases
where the MRI is also active.

\begin{acknowledgments}
This research was supported in part by the
Swedish Research Council (Vetenskapsr{\aa}det) under grant No.\ 2019-04234,
the National Science Foundation under grants No.\ NSF PHY-2309135,
AST-2307698, and AST-2408411, and NASA grant No.\ 80NSSC22K0825.
We acknowledge the allocation of computing resources provided by the
Swedish National Allocations Committee at the Center for
Parallel Computers at the Royal Institute of Technology in Stockholm.
The authors wish to thank the participants of the Nordita program ``Stellar
Convection: Modelling, Theory, and Observations'' in August and September
2024 and the Munich Institute for Astro, Particle and BioPhysics
(MIAPbP), which is funded by the Deutsche Forschungsgemeinschaft (DFG,
German Research Foundation) under Germany's Excellence Strategy --
EXC-2094 -- 390783311, for many inspiring discussions.

\vspace{2mm}\noindent
{\em Software and Data Availability.}
The source code used for the simulations of this study,
the {\sc Pencil Code} \citep{PC}, is freely available on
\url{https://github.com/pencil-code}.
The simulation setups and corresponding input
and reduced output data are freely available on
\dataset[http://doi.org/10.5281/zenodo.15258044]{http://doi.org/10.5281/zenodo.15258044}.
\end{acknowledgments}

\bibliographystyle{aasjournal}
\bibliography{ref}

\end{document}